\documentclass{aa}

\usepackage{graphicx}
\usepackage{txfonts}
\usepackage{natbib}
\usepackage{hyperref}
\usepackage{bm}
\AtBeginDocument{\renewcommand{\d}{\textrm{d}}}
\AtBeginDocument{\renewcommand{\i}{\textrm{i}}}

\begin{document} 

\title{Rotation of an oblate satellite: Chaos control}

        \author{M. Tarnopolski
               \inst{1}
        }

        \institute{Astronomical Observatory, Jagiellonian University,
                Orla 171, 30-244 Krak\'ow, Poland\\
                \email{mariusz.tarnopolski@uj.edu.pl}
        }

        \date{Received; accepted}

\abstract
{}
{This paper investigates the chaotic rotation of an oblate satellite in the context of chaos control.}
{A model of planar oscillations, described with the Beletskii equation, was investigated. The Hamiltonian formalism was utilized to employ a control method for suppressing chaos.}
{An additive control term, which is an order of magnitude smaller than the potential, is constructed. This allows not only for significantly diminished diffusion of the trajectory in the phase space, but turns the purely chaotic motion into strictly periodic motion.}
{}

\keywords{celestial mechanics -- chaos -- methods: numerical}
\titlerunning{Rotation of an oblate satellite: Chaos control}
\maketitle

\section{Introduction}\label{sect1}

After Voyager~2 \citep{smith} obtained high-quality images of Hyperion \citep{bond,lassel}, the highly aspherical moon of Saturn, it became clear that it remained in an exotic rotational state \citep{klav1,klav2,black,devyatkin,hicks,thomas10,harbison}. In a seminal paper, \citet{wisdom} predicted that Hyperion rotates chaotically, analyzed the phase space of a model of planar rotation, showed that it has an unstable attitude, and computed the Lyapunov time to be about 10 times the orbital period (i.e. $10\times 21.28\,{\rm d}$). Since then, the analyses of chaotic rotation of an oblate celestial body has become very common, regarding Hyperion, per se, as well as other solar system satellites.

For instance, \citet{boyd} employed the method of close returns to a sparse data set of dynamical states of Hyperion simulated with Euler equations, and found that a time series spanning about 2.6 years of data is sufficient to infer the temporary rotational state (chaotic/regular). These findings agree with the recent results of \citet{tarnopolski}, who showed that to extract a maximal Lyapunov exponent from ground-based photometric observations of Hyperion's light curve, at least one year of well-sampled data is required, but a three-year data set would be desirable. \citet{black} performed numerical simulations with the full set of Euler equations to model the long-term dynamical evolution, and found that the chaotic tumbling of Hyperion leads to transitions between temporarily regular and chaotic rotation with a period of hundreds of days up to thousands of years. It was also shown that the Voyager~2 images of Hyperion indicate that the motion was predictable at the time of the passage \citep{thomas95}.

\citet{beletskii} examined a number of theoretical models, including the gravitational, magnetic, and tidal moments and analyzed the rotation in the gravitational field of two centers. The structure of the respective phase spaces was investigated with Poincar\'e surfaces of section. The stability of spin-orbit resonances, with application to the solar system satellites, was inferred based on a series expansion of the equation of planar rotation (\citealt{celletti1,celletti2}; see also \citealt{celletti}). The Lyapunov spectra were exhaustively examined for several satellites \citep{shevchenko1,shevchenko,kouprianov1,kouprianov2}, and the Lyapunov times spanned 1.5--7 orbital periods for Hyperion. An interesting possibility that Enceladus might be locked in a 1:3 librational-orbital secondary resonance was investigated using the model of planar rotation within the Hamiltonian formalism \citep{wisdom2}. The dynamical stability was examined for a number of known satellites by \citet{melnikov}; in particular, the synchronous spin-orbit resonance in case of Hyperion was confirmed to be unstable. The Hamiltonian formalism has also been employed in the research of secondary resonances \citep{gkolias}, and has taken non-conservative forces into account \citep{gkolias2}. Finally, regarding the influence of a secondary body on the rotation of an oblate moon, \citet{tarnopolski2} showed, using the correlation dimension and bifurcation diagrams, that the introduction of an additional satellite can change the rotation from regular to chaotic.

Well-defined methods to reduce chaos in physical systems have been known for a long time now (e.g. \citealt{ott,pyragas}). In general, a chaos control scheme already demonstrated its usefulness in astrodynamical applications, for example in maneuvering the ISEE-3/ICE-3 satellite to reach the Giacobini-Zinner comet in 1985 \citep{shinbrot}. Regarding rotation of an oblate moon, the full set of modified Euler equations was investigated numerically with various methods in the context of chaos control \citep{tsui} for a satellite with thrusters. Investigation of the Mimas-Tethys system was successfully conducted by means of a Hamiltonian chaos control method \citep{khan2}. While still rather futuristic, an ability to construct efficient control terms might prove to be important in future asteroid capture missions and mining attempts \citep{kargel,sonter,koon,leva,elvis}.

The knowledge about the rotational state of a celestial body proved to be crucial in the 2011 comet Tempel 1 flyby of Stardust-NExT \citep{veverka}. Tempel 1 was the target of the Deep Impact mission in 2005 \citep{ahearn}. The mission's aim was to make the impactor collide with the comet to excavate a crater to allow investigation of its interior structure. The crater was not measured directly after the impact owing to a large cloud of dust that obscured the view of the orbiting spacecraft. Hence one of the objectives of Stardust-NExT was to image the site of the impactor's collision with the object. The rotational state of the comet needed to be accurately predicted so that the site of interest would be visible to the spacecraft and well illuminated during the flyby. The rotational period was shown to be decreasing \citep{belton} and the time of arrival to the comet was adjusted by $8.5\,{\rm h}$ one year prior to the encounter \citep{veverka}. While, unlike Hyperion, Tempel 1 is not rotating chaotically, it is a plain illustration of the importance of the rotational state of small solar system bodies, among which chaotic rotation is expected to be common \citep{jacobson2,jafari}.

Herein, a construction of a control term that reduces chaos substantially, down to a strictly periodic rotation, is presented. Using the \citet{beletskii3} equation, the Hamiltonian setting of the problem is employed to investigate the prospects of chaos control within the framework of \citet{ciraolo}. Numerical examples are carried out with the parameters suitable for Hyperion, but the approach is general enough to be widely applicable and not restricted to only some ranges of parameter values; the latter is also illustrated with parameters typical for solar system asteroids.

This paper is organized in the following manner. In Sect.~\ref{sect2} the rotational model is introduced, and analytical solutions in some specific cases are presented for completeness. In Sect.~\ref{hamiltonian} the Hamiltonian approach is discussed and the chaos control method is described. Sect.~\ref{sect262} shows that the method is able to suppress diffusion of a chaotic trajectory in the phase space effectively. Discussion and conclusions are gathered in Sect.~\ref{sect5}. The computer algebra system \textsc{Mathematica} is applied throughout this paper.

\section{Rotational model of an oblate moon}\label{sect2}

\subsection{Equations of motion}\label{sect21}

\begin{figure}
\centering
\includegraphics[width=0.5\textwidth]{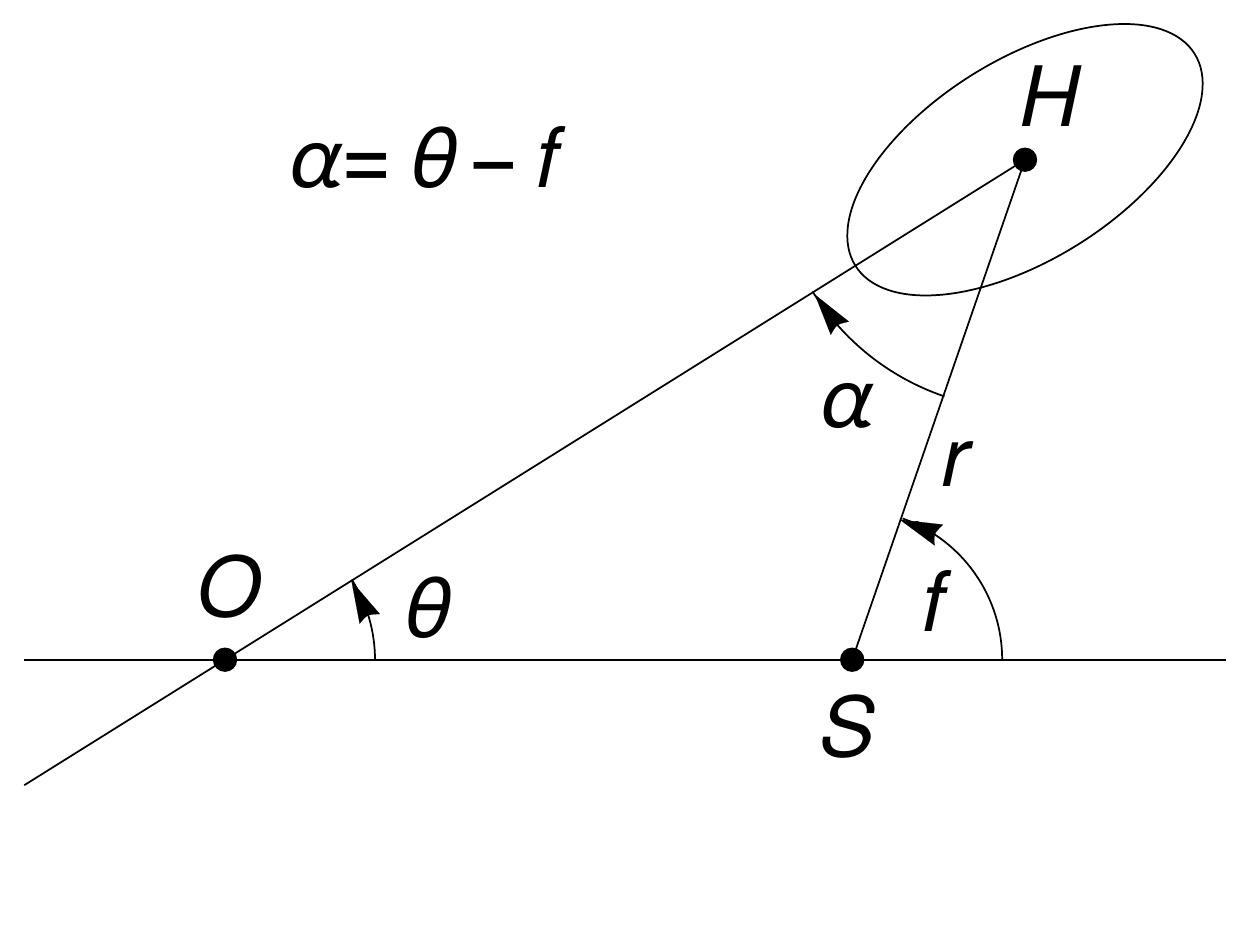}
\caption{Rotational model of an oblate moon ($H$) orbiting around a central body (denoted with $S$); $\alpha$ is the angle between the long axis of the moon and the direction to the planet, and $\theta$ is the dynamical angle from Eq.~(\ref{eq3}) and (\ref{eq5}), where $f$ is the true anomaly.}
\label{fig1}
\end{figure}

The rotational equation of motion can be derived based on the following assumptions \citep{danby,goldreich,wisdom,sussman,greiner}:
\begin{enumerate}
\item The orbit of the satellite around its host planet is Keplerian with eccentricity $e$, i.e. the distance $r$ between the satellite and its host planet is (with major semi-axis $a=1$)
\begin{equation}
r=\frac{\left(1-e^2\right)}{1+e\cos f},
\label{eq1}
\end{equation}
where $f$ is the true anomaly given by
\begin{equation}
\dot{f}=\frac{\left(1+e\cos f\right)^2}{\left(1-e^2\right)^{3/2}},
\label{eq2}
\end{equation}
where the overdot denotes differentiation with respect to time (see Fig.~\ref{fig1} for the geometrical setting). The units are chosen so that the orbital period $T=2\pi$, where $a=1$ leads to the orbital mean motion $n=1$.
\item The satellite is modeled as a triaxial ellipsoid, with principal moments of inertia $A<B<C$.
\item The spin axis is fixed and perpendicular to the orbit plane; it is aligned with the shortest physical axis (i.e. the axis related to the largest principal moment of inertia $C$).
\end{enumerate}
With the oblateness defined as $\omega^2=\frac{3(B-A)}{C}$, the equation of motion takes the form
\begin{equation}
\ddot{\theta}+\frac{\omega^2}{2r^3}\sin 2\left(\theta-f\right)=0,
\label{eq3}
\end{equation}
where $\theta$ is the angular orientation of the satellite relative to some arbitrary line in space (see Fig.~\ref{fig1}).

It should be emphasized that while the last assumption is valid for most solar system satellites, as the angular momentum is assumed to be constant, it was shown that the chaotic state is attitude unstable in case of Hyperion \citep{wisdom}. Therefore, the model investigated herein should be thought of as an illustrative first approximation. However, a more astrodynamically realistic setting is also examined further on.

Using the well-known general relations in the two-body motion, where $M$ is the mass of the planet, $\mu=GM$, $h=r^2\dot{f}$, $h^2=\mu l$, $l/r=1+e\cos f$ [i.e. Eq.~(\ref{eq1})], $l=a(1-e^2)$ \citep{khan}, one can transform Eq.~(\ref{eq3}) so that $f$ is the independent variable [the famous \citet{beletskii3} equation; see Appendix \ref{appA}],
\begin{equation}
(1+e\cos f)\frac{\d^2\theta}{\d f^2}-2e\sin f\frac{\d \theta}{\d f}-\frac{\omega^2}{2}\sin 2(f-\theta)=0,
\label{eq5}
\end{equation}
which as a dynamical system $\dot{\vec{x}}=\vec{F}(\vec{x})$ reads
\begin{subequations}
\begin{align}
\theta' & = \eta, \label{eq6a} \\
\eta'   & = \frac{2e\sin f\cdot\eta-\frac{\omega^2}{2}\sin 2(\theta-f)}{1+e\cos f}, \label{eq6b}
\end{align}
\label{eq6}%
\end{subequations}
where the prime denotes differentiation with respect to $f$. This system is non-autonomous.\footnote{Writing Eq.~(\ref{eq2}) and (\ref{eq3}) together as a system of three autonomous first-order differential equations yields a so-called 1.5 degree of freedom system; however, only two equations---those from a second-order Eq.~(\ref{eq3})---carry dynamical information.}

It should be stressed that the equation of motion in the temporal domain, i.e. Eq.~(\ref{eq3}), can also be obtained from a Hamiltonian \citep{sussman,flynn,celletti4} describing the rotational motion, with an auxiliary differential equation governing the evolution of the true anomaly, Eq.~(\ref{eq2}), as a constraint coming from the description of the orbit. Such a Hamiltonian describes a 1-degree of freedom, non-autonomous system with implicit time dependence through $2\pi$-periodic functions $f=f(t)$ and $r=r(t)$.

\subsection{Analytical instances}\label{sect22}

For $e=0$ and $\omega^2=0$, i.e. for a circular orbit and spherical symmetry of the satellite, Eq.~(\ref{eq3}) leads to a uniform rotation, $\theta(t)=\dot{\theta}_0t+\theta_0$.

When $e=0$ and $\omega^2\neq 0$, Eq.~(\ref{eq2}) is solved trivially by $f(t)=t+f_0$, and Eq.~(\ref{eq3}) becomes the pendulum equation,
\begin{equation}
\ddot{\alpha}+\frac{\omega^2}{2}\sin 2\alpha=0,
\label{eq7}
\end{equation}
where $\alpha=\theta-f$ (see Fig.~\ref{fig1}), which is solved by the Jacobi elliptic functions \citep{lowenstein,lara}. With $\psi=2\alpha$,
\begin{equation}
\dot{\psi}=\pm\sqrt{2\left(4\mathcal{E}+\omega^2\cos\psi\right)},
\label{eq8}
\end{equation}
where $\mathcal{E}=\frac{\ddot{\alpha}}{2}-\frac{\omega^2}{4}\cos 2\alpha$ is a constant of motion. In Eq.~(\ref{eq8}), $4\mathcal{E}$ is the total energy of the pendulum, while $\omega^2$ is the maximal value of the potential energy. For $\mathcal{E}<\frac{\omega^2}{4}$, the motion is a libration in the orbital plane, and it is a rotation for $\mathcal{E}>\frac{\omega^2}{4}$ \citep{murray}. For the border case, $\mathcal{E}=\frac{\omega^2}{4}$, the motion takes place on a separatrix, with an initial condition $\alpha(0)=0$, i.e.\begin{equation}
\psi(t)=4\arctan\left({\rm e}^{\pm\omega t}\right)-\pi,
\label{eq9}
\end{equation}
which is an unstable trajectory. The separatrix consists of two branches that form a cross in the phase space; this point is an unstable stationary point $\left(\pm\pi,0\right)$. Asymptotically, every initial condition ends in this point.

\begin{figure*}
\includegraphics[width=\textwidth]{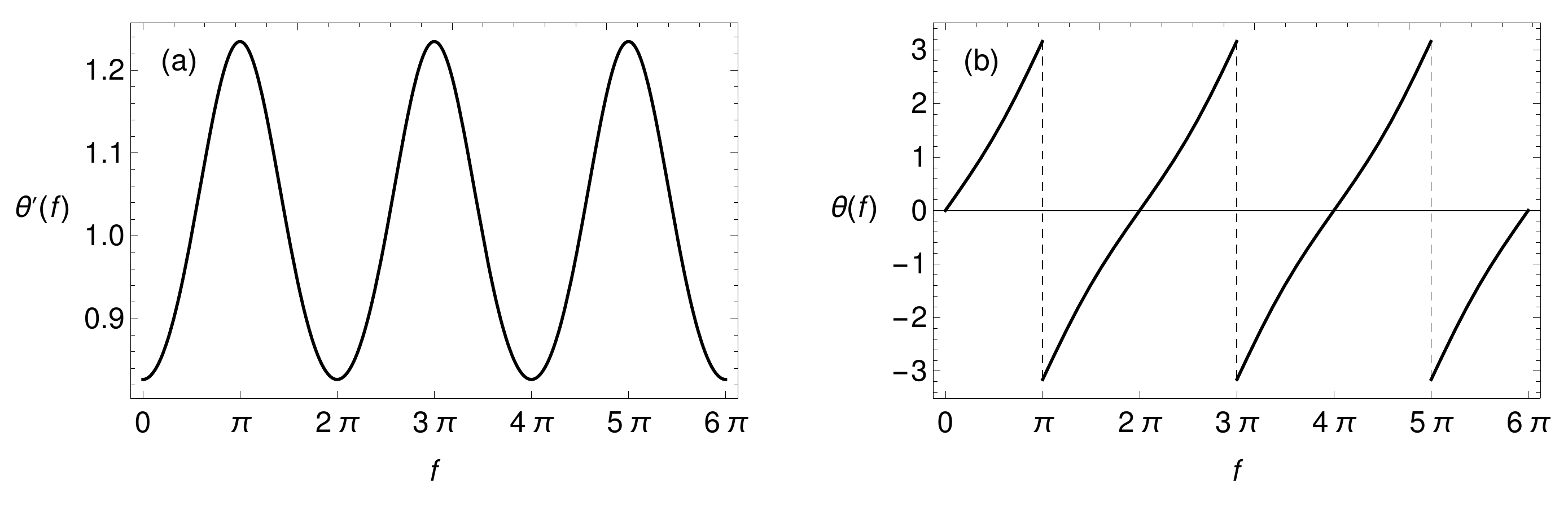}
\caption{Solution of Eq.~(\ref{eq10}). Dependence of (a) the angular velocity $\theta'(f)$ and (b) the orientation $\theta(f)$ on the true anomaly $f$. Constants of integration are $C_1=1$, $C_2=0$.}
\label{fig2}
\end{figure*}

The last simple case is $\omega^2=0$ and $e\neq 0$. This again leads, via Eq.~(\ref{eq3}), to $\theta(t)=\dot{\theta}_0t+\theta_0$. But now, Eq.~(\ref{eq5}) takes the form
\begin{equation}
\left(1+e\cos f\right)\theta''(f)-2e\theta'(f)\sin f=0,
\label{eq10}
\end{equation}
which allows us to investigate how the rotation of a satellite depends on its orbital location. Eq.~(\ref{eq10}) is integrable and yields the angular velocity
\begin{equation}
\theta'(f)=\frac{C_1}{\left(1+e\cos f\right)^2},
\label{eq11}
\end{equation}
which in turn can be also directly integrated (see Fig.~\ref{fig2}),
\begin{equation}
\begin{split}
& \theta(f)=C_1\Bigg[\frac{e\sin f}{\left(e^2-1\right)\left(1+e\cos f\right)} + \\
& \hspace{1.5cm} \frac{2}{\left(\sqrt{1-e^2}\right)^{3/2}}\arctan\frac{\left(1-e\right)\tan\frac{f}{2}}{\sqrt{1-e^2}}\Bigg]+C_2,
\label{eq12}
\end{split}
\end{equation}
where $C_1$ and $C_2$ are constants of integration.

\subsection{Fourier series expansion}\label{sect23}

Eq.~(\ref{eq3}) can be expanded into a Fourier-like series, formally valid for all $\omega^2$,
\begin{equation}
\ddot{\theta}+\frac{\omega^2}{2}\sum\limits_{k=-\infty}^{\infty}H\left(\frac{k}{2},e\right)\sin (2\theta-kt)=0,
\label{eq13}
\end{equation}
which can be naturally averaged over one orbital period to give
\begin{equation}
\ddot{\gamma}+\frac{\omega^2}{2}H\left(\frac{k}{2},e\right)\sin 2\gamma=0,
\label{eq14}
\end{equation}
i.e. the pendulum equation from Eq.~(\ref{eq7}), with $\omega\rightarrow\omega\sqrt{|H(k/2,e)|}$, to which the discussion from Sect.~\ref{sect22} applies. Here, $\gamma=\theta-\frac{k}{2}t$ is a resonant variable.

The coefficients $H(k/2,e)$ were calculated by \citet{cayley}, and the first few of these coefficients can be found tabulated in \citep{goldreich,wisdom,celletti1,murray}. For a set $k$, $H(k/2,e)$ is a power series in $e$, with its dominant term $\propto e^{|k-2|}$. Thence, higher order terms are negligible in most cases (see also \citealt{celletti2}). The coefficients are given by an integral formula \citep{murray,beletskii4}
\begin{equation}
H\left(\frac{k}{2},e\right)=\frac{1}{\pi (1-e^2)^{3/2}}\int\limits_0^\pi(1+e\cos f)[k\tau(f)-2f]df,
\label{eq15}
\end{equation}
where
\begin{equation}
\tau(f)=2\arctan\sqrt{\frac{1-e}{1+e}}\tan\frac{f}{2}-\frac{e\sqrt{1-e^2}}{1+e\cos f}\sin f.
\label{eq16}
\end{equation}
Truncations of the series allow  us to infer the stability of the spin-orbit resonances in the solar system \citep{celletti2}. These resonances occur when $\dot{\gamma}\approx 0$, i.e. $|\dot{\theta}-\frac{k}{2}|\ll \frac{1}{2}$. Locations in the phase space of the main resonances can be obtained by an analysis of Eq.~(\ref{eq3}) or (\ref{eq5}) and are given in \citep{wisdom} and \citep{black}.

Based on the \citet{chirikov} criterion (see also \citealt{lichtenberg}), the onset of chaos is predicted to occur when the 1:1 and 3:2 resonances overlap, which happens when the critical value of $\omega$ is $\omega^R=1/\left(2+\sqrt{14e}\right)$. For Hyperion's eccentricity, $e=0.1$, $\omega^R\approx 0.31$. This was confirmed with numerical simulations \citep{wisdom,tarnopolski2}.

\section{Hamiltonian formalism and chaos control}\label{hamiltonian}

\subsection{Hamiltonian}\label{sect24}

Using the Euler-Lagrange equations $\frac{\d}{\d f}\frac{\partial\mathcal{L}}{\partial\theta'}=\frac{\partial\mathcal{L}}{\partial\theta}$ on a general Lagrangian
\begin{equation}
\mathcal{L}=\frac{1}{2}G\left(\theta(f),f\right)\theta'(f)^2+F\left(\theta(f),f\right)\theta'(f)-\mathcal{V}\left(\theta(f),f\right),
\label{eq17}
\end{equation}
one finds that Eq.~(\ref{eq5}) is obtained for
\begin{equation}
\mathcal{L}(\theta,\theta')=\frac{1}{2}\left(1+e\cos f\right)^2\theta'^2+\frac{\omega^2}{4}\left(1+e\cos f\right)\cos 2(f-\theta).
\label{eq18}
\end{equation}
With $p=\frac{\partial\mathcal{L}}{\partial\theta'}=\left(1+e\cos f\right)^2 \theta'$, the Hamiltonian can be obtained through the Legendre transformation $\mathcal{H}=p\theta'-\mathcal{L}$ as
\begin{equation}
\mathcal{H}(\theta,p)=\frac{p^2}{2\left(1+e\cos f\right)^2}-\frac{\omega^2}{4}\left(1+e\cos f\right)\cos 2(f-\theta).
\label{eq19}
\end{equation}
The Hamiltonian $\mathcal{H}$ is explicitly dependent on $f$, which has the meaning of time in this setting, hence it is not a constant of motion. In particular, when $\omega^2=0$, the Hamilton equations lead to $p\in {\rm const.}$ and to Eq.~(\ref{eq11}) with $C_1=p$, as should be expected.

\subsection{Phase space volume conservation}\label{sect25}

Consider a generic Hamiltonian $\mathcal{H}(q,p)$. The Liouville theorem states that the flow is conservative, $\nabla\cdot\vec{F}=0$, i.e. the volume is preserved as the system evolves with time. Now consider a general dynamical system $\dot{\vec{x}}=\vec{F}(\vec{x})$. A subset of the phase space with initial volume $V_0$ evolves according to the equation
\begin{equation}
\frac{\d}{\d t}\ln V(t)=\nabla\cdot\vec{F}.
\label{eq20}
\end{equation}
If the system under consideration is Hamiltonian, then this relation reverts to the Liouville theorem via the Hamilton equations $\dot{q}=\frac{\partial\mathcal{H}}{\partial p}$, $\dot{p}=-\frac{\partial\mathcal{H}}{\partial q}$. Thence, the flow in the $(\theta,p)$ phase space described by the Hamiltonian in Eq.~(\ref{eq19}) is obviously conservative. 

Consider, however, the dynamical system in Eq.~(\ref{eq6}). Its phase space is $(\theta,\theta')$, and
\begin{equation}
\frac{\d}{\d f}\ln V(f) = \nabla\cdot\vec{F} = \frac{2e\sin f}{1+e\cos f}.
\label{eq21}
\end{equation}
This equation can be solved for $V(f)$ as follows:
\begin{equation}
V(f)=\frac{V_0(1+e)^2}{(1+e\cos f)^2}
\label{eq22}
,\end{equation}
where $V_0=V(0)$. This function is $2\pi$ periodic.

Owing to Eq.~(\ref{eq21}) the flow is {\it not} conservative for the system in Eq.~(\ref{eq6}). Indeed, the flow is conservative in canonical coordinates $(\theta,p)$ with $p=\left(1+e\cos f\right)^2 \theta'$ (see Sect.~\ref{sect24}). The volume does not need to be conserved in other coordinates. In general, consider a rectangular area in a two-dimensional phase space $(q,p)$: $V_0=q_0p_0=q_tp_t=V_t\equiv V(t)$. The relation between $p$ and $\dot{q}$ is $\dot{q}=\frac{p}{\zeta(t)^2}$, where $\zeta(t)$ is a scalar function. Then, the corresponding volume in the $(q,\dot{q})$ space is $\tilde{V}_0=\frac{q_0p_0}{\zeta(0)^2}$, and $\tilde{V}_t=\frac{q_tp_t}{\zeta(t)^2}$. $\tilde{V}_t\neq \tilde{V}_0$, but $\tilde{V}_t\zeta(t)^2=\tilde{V}_0\zeta(0)^2\,\Rightarrow\,\tilde{V}_t=\frac{\tilde{V}_0\zeta(0)^2}{\zeta(t)^2}$, which is the form of Eq.~(\ref{eq22}).

\subsection{Chaos control method}\label{sect26}

The method of \citet{ciraolo} is employed and briefly described as follows. Consider an autonomous Hamiltonian $\mathcal{H}=\mathcal{H}_0+\varepsilon \mathcal{V}$ with a $2n$-dimensional phase space, where $\mathcal{H}_0$ is the Hamiltonian of an integrable system and $\varepsilon$ is a multiplicative parameter (relatively small). The aim is to find a control term $\mathcal{F}$ of order $O(\varepsilon^2)$ such that the motion described by a new Hamiltonian with the control term $\tilde{\mathcal{H}}=\mathcal{H}+\mathcal{F}=\mathcal{H}_0+\varepsilon \mathcal{V}+\mathcal{F}$ is much less diffused in the phase space, i.e. it is much more ordered. This is accomplished in the next steps.

First, with $\vec{A}$ being the actions and $\bm{\varphi}$ the angles, $\mathcal{V}$ is decomposed as
\begin{equation}
\mathcal{V}=\sum\limits_{\vec{k}\in\mathbb{Z}^n}\mathcal{V}_{\vec{k}}\exp{(\i\vec{k}\cdot\bm{\varphi})}.
\label{eq28}
\end{equation}
For this purpose, integers $\vec{k}=(k_1,k_2,\ldots,k_n)$ need to be found. Next, the frequency vector is constructed, $\bm{\xi}=\frac{\partial\mathcal{H}_0}{\partial\vec{A}}$. With $\mathcal{V}$ and $\bm{\xi}$, one defines the following operators via their action on $\mathcal{V}$:
\begin{subequations}
\begin{align}
\Gamma \mathcal{V} = \sum_{\substack{\vec{k}\in\mathbb{Z}^n\\ \bm{\xi}\cdot\vec{k}\neq 0}}\frac{\mathcal{V}_{\vec{k}}}{\i \bm{\xi}\cdot\vec{k}}\exp{(\i \vec{k}\cdot\bm{\varphi})}, \label{eq29a} \\
\mathcal{R}\mathcal{V} = \sum_{\substack{\vec{k}\in\mathbb{Z}^n\\ \bm{\xi}\cdot\vec{k}=0}} \mathcal{V}_{\vec{k}}\exp{(\i \vec{k}\cdot\bm{\varphi})}, \label{eq29b} \\
\mathcal{N}\mathcal{V} = \sum_{\substack{\vec{k}\in\mathbb{Z}^n\\ \bm{\xi}\cdot\vec{k}\neq 0}} \mathcal{V}_{\vec{k}}\exp{(\i \vec{k}\cdot\bm{\varphi})}, \label{eq29c}
\end{align}
\label{eq29}%
\end{subequations}
where $\mathcal{R}\mathcal{V}$ is the resonant and $\mathcal{N}\mathcal{V}$ is the non-resonant part of $\mathcal{V}$.

For the purpose of this work, it is sufficient to describe the construction of $\mathcal{F}$ in the case when $\mathcal{R}\mathcal{V}=0$. The control term is given by a series
\begin{equation}
\mathcal{F}=\sum\limits_{s=2}^{\infty}\mathcal{F}_s,
\label{eq30}
\end{equation}
where
\begin{equation}
\mathcal{F}_s=-\frac{1}{s}\left\{\Gamma \mathcal{V},\mathcal{V}\right\}
\label{eq31}
\end{equation}
where $\mathcal{F}_1=\mathcal{V}$ and $\left\{\cdot,\cdot\right\}$ is the Poisson bracket. The first term of the series in Eq.~(\ref{eq30}) takes the form
\begin{equation}
\mathcal{F}_2=-\frac{1}{2}\left\{\Gamma \mathcal{V},\mathcal{V}\right\}.
\label{eq32}
\end{equation}
The whole infinite series does not need to be calculated. In fact, truncating the series after the first term turns out to effectively suppress chaos. To additionally improve the performance, a scaling factor $\eta$ is introduced multiplicatively, i.e. $\mathcal{F}_2\rightarrow\eta\mathcal{F}_2$, and its optimal value is found numerically.

\section{Results}\label{sect262}

Consider the Hamiltonian from Eq.~(\ref{eq19}). It has 1.5 degrees of freedom, so one needs to introduce a second action, $E$, to make it autonomous. It takes then the form
\begin{equation}
\mathcal{H}=\underbrace{\frac{p^2}{2\left(1+e\cos f\right)^2}+E}_{\mathcal{H}_0} \underbrace{-\frac{\omega^2}{4}\left(1+e\cos f\right)\cos 2(\theta-f)}_{\mathcal{V}}.
\label{eq33}
\end{equation}
One denotes $\vec{A}=(p,E)$, $\bm{\varphi}=(\theta,f)$. The frequency vector is $\bm{\xi}=\left(\frac{\partial\mathcal{H}_0}{\partial p},\frac{\partial\mathcal{H}_0}{\partial E}\right)=\left(\frac{p}{(1+e\cos f)^2},1\right)$. The potential $\mathcal{V}$ can be decomposed into the form in Eq.~(\ref{eq28}) with
$$\vec{k}=(k_1,k_2)\in\{(-2,1),(-2,2),(-2,3),(2,-1),(2,-2),(2,-3)\},$$
and $\mathcal{V}_{\vec{k}}$ are simple monomials in $e$ and $\omega$. Thence, for example $\bm{\xi}\cdot(-2,1)=1-\frac{2p}{(1+e\cos f)^2}$. Overall, $\bm{\xi}\cdot\vec{k}\neq 0$ for $p\neq c(1+e\cos f)^2\theta'$, $c=\frac{1}{2},1,\frac{3}{2}$, i.e. $\theta'\neq \frac{1}{2},1,\frac{3}{2}$, which are major spin-orbit resonances. It is assumed hereinafter that the trajectory is far from these resonances, which means that $\mathcal{R}\mathcal{V}=0$, $\mathcal{N}\mathcal{V}=\mathcal{V}$. Eq.~(\ref{eq29a}) gives a lengthy output, as does $\mathcal{F}_2$ in terms of $p$ (see Appendix \ref{appB} for explicit formulae). With the substitution $p=(1+e\cos f)^2\theta'$ one gets a much more compact expression, which expanded into a zeroth-order series takes the final form,
\begin{equation}
\begin{split}
& \mathcal{F}_{2(0)}=-\frac{1}{288 (1+e\cos f)}\Bigg\{ \omega ^4 \sin 2(f-\theta ) \Big\{2 e \big[ 9 \sin (f-2 \theta )+ \\
& \hspace{1.8cm} + \sin (3 f-2 \theta ) \big] + 9 \sin 2 (f-\theta )\Big\} \Bigg\},
\label{eq34}
\end{split}
\end{equation}
which is indeed $O(\omega^4)$, and the index in brackets emphasizes it is a zeroth-order expansion of $\mathcal{F}_2$.

The Hamiltonian with the control term is
\begin{equation}
\tilde{\mathcal{H}}=\frac{p^2}{2(1+e\cos f)^2}-\frac{\omega^2}{4}(1+e\cos f)\cos 2(\theta-f)+\eta\mathcal{F}_{2(0)},
\label{eq35}
\end{equation}
where the scaling factor $\eta$ was already introduced (compare with Sect.~\ref{sect26}). Eventually the Hamilton equations yield
\begin{equation}
\begin{split}
& (1 + e \cos f)\theta'' - 2 e \sin f \theta' = \\
& \frac{1}{72 (1 + e \cos f)^2} \Bigg\{9 \omega^2 \sin 2 (f - \theta ) \big[2 e^2 (1 + \cos 2 f ) + 8 e \cos f + \\
& - \eta \omega^2 \cos 2 (f - \theta ) +  4 \big] - \eta e \omega^4 \big[9 \sin (3 f - 4 \theta ) + \sin (5 f - 4 \theta )\big]\Bigg\}.
\end{split}
\label{eq36}
\end{equation}

\begin{figure*}
\includegraphics[width=\textwidth]{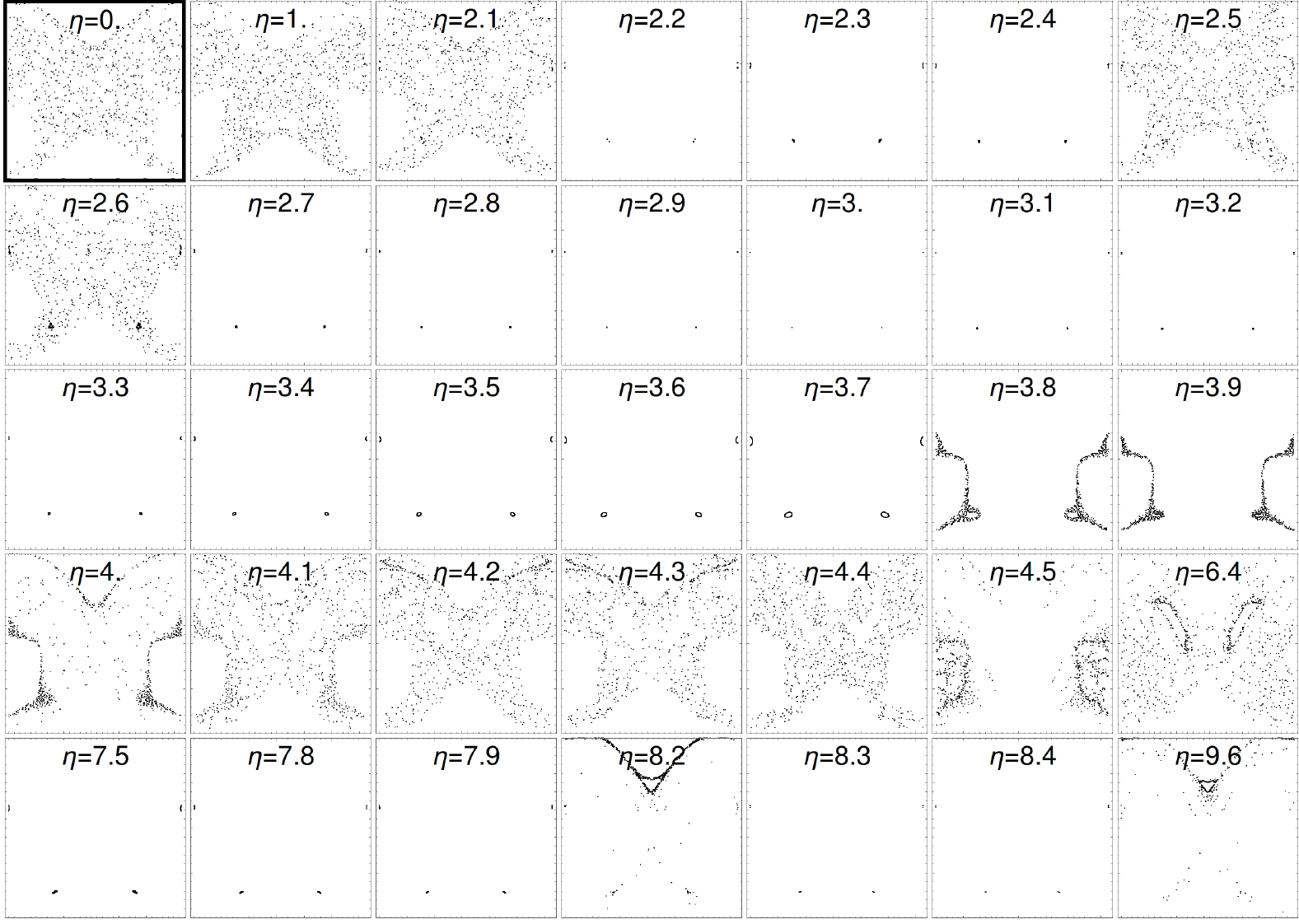}
\caption{Poincar\'e surfaces of section obtained by integrating Eq.~(\ref{eq36}) and recording the points $\left(\theta(f),\theta'(f)\right)$ with a step of $\Delta f=2\pi$. The highlighted plot corresponds to $\eta=0$, i.e. no control term. The most prominent suppression of chaos is obtained with $\eta=3$.}
\label{fig14}
\end{figure*}

\begin{figure}
\centering
\includegraphics[width=0.5\textwidth]{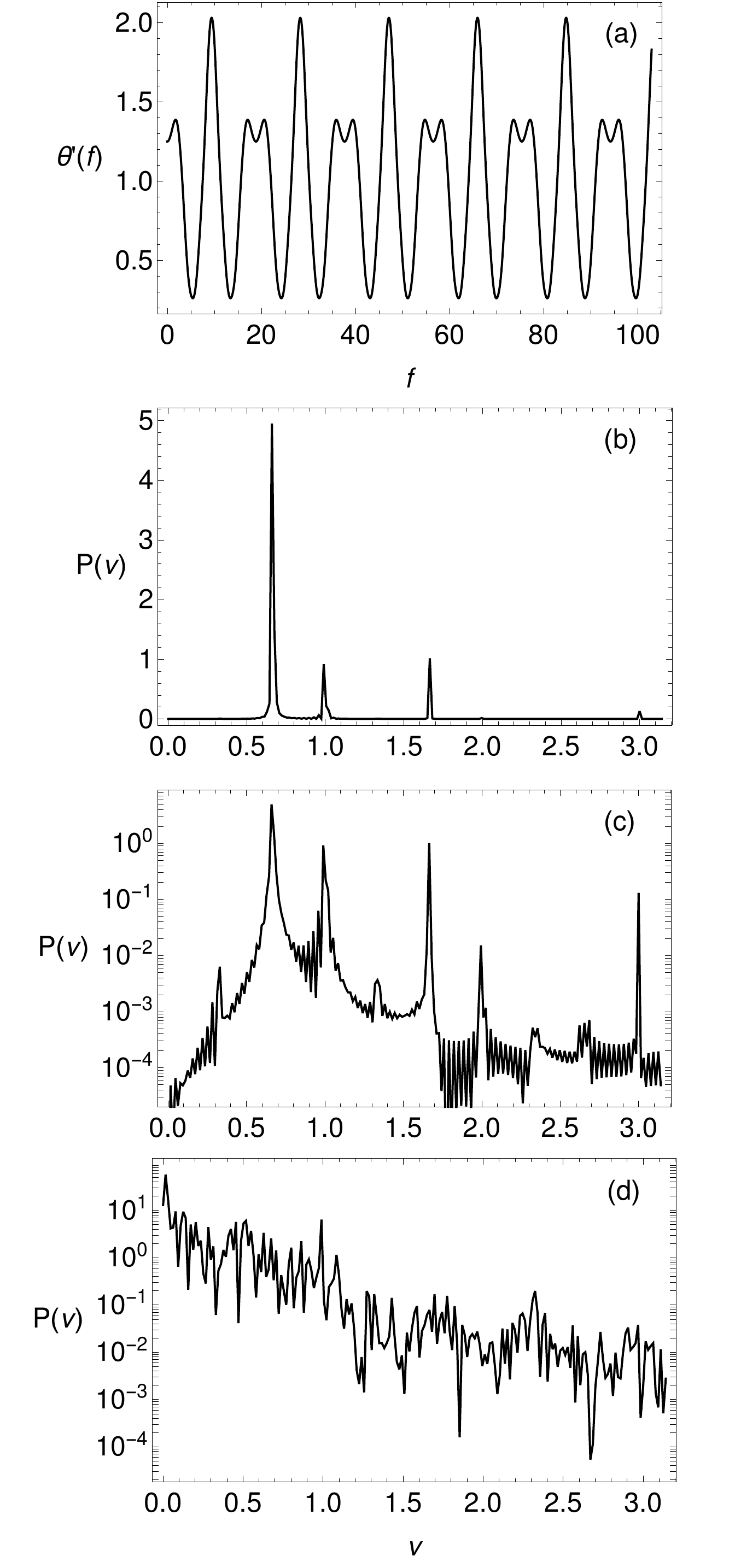}
\caption{(a) Solution of Eq.~(\ref{eq36}) for the suppressed case $\eta=3$. (b) Its power spectrum in a linear scale and (c) in log scale. (d) A broadband spectrum for the chaotic case $\eta=0$.}
\label{fig15}
\end{figure}

Eq.~(\ref{eq36}) was integrated numerically for $e=0.1$, $\omega=0.89$, $f\in[0,5000]$, and $\eta\in[0,9.9]$ with a step of 0.1. The case $\eta=0$ corresponds to the lack of a control term, i.e. the Hamiltonian in Eq.~(\ref{eq19}) yielding Eq.~(\ref{eq5}) as the equation of motion, hence serves as a reference. Fig.~\ref{fig14} shows the obtained Poincar\'e surfaces of section.
The plot corresponding to the lack of control term is highlighted with the thick frame. All plots up to $\eta=2.1$ do not qualitatively influence the surface of section, i.e. the diffusion was not diminished at all. Impressively, $\eta=2.2\!-\!2.4$ suppresses chaos to a very small region of the phase space, forming closed curves. At a slightly higher $\eta$, chaotic motion bursts out, to be suppressed again for a wide range of the scaling factor, $\eta=2.7\!-\!3.7$. The most prominent reduction of the chaotic behavior is observed for $\eta=3$, where the motion is turned to periodic. To emphasize the extent of this reduction, Fig.~\ref{fig15} shows a power spectrum (a Lomb-Scargle periodogram; \citealt{lomb,scargle}) of the $\theta'$ time series; the case $\eta=0$ (chaotic) is displayed for comparison. When $\eta$ is increased further, the motion turns back to chaotic with occasional signs of stickiness, for example $\eta=4.5$ or 6.4. The spectacular suppression occurs again for $\eta=7.5,7.8,7.9,8.3,8.4$; on the other hand, stickiness is manifested for $\eta=8.2$ and 9.6. The stickiness occurring for these higher values of $\eta$ is different than for such lower values: for $\eta=4.5$ the points gather around the 1:1 spin-orbit resonance (compare with \citealt{black}), while for $\eta=8.2$ the gathering happens near the upper separatrix. The maximal value of the potential $\mathcal{V}$ is 0.218, while for the control term $\mathcal{F}_{2(0)}$ (i.e. taking $\eta=1$) it does not exceed 0.022; i.e. the control term is only about 10\% of the potential. The value $\eta\gtrsim 10$ is useless since the control term is then of the same order as the potential $\mathcal{V}$. However, for $\eta\sim 2.2$ (compare with Fig.~\ref{fig14}) it is still nearly an order of magnitude smaller and is sufficient to suppress the chaotic behavior of the investigated system almost entirely. The range $\eta\in[0,2.2]$ was also sampled with a smaller step of 0.01, but the resulting surfaces of section did not reveal any signs of chaos suppression. Also the first-order series expansion of $\mathcal{F}_2$, denoted $\mathcal{F}_{2(1)}$ (not shown), was tested; it allowed chaos to be suppressed down to a periodic motion with $\eta=0.7$, but its overall maximal value attained 1/3 of the maximal value of the potential $\mathcal{V}$, i.e. there was no improvement compared to the performance of $\mathcal{F}_{2(0)}$.

Finally, to examine a more astrodynamically realistic scenario, with $\omega^2=0.25$ and $e=0.005$ the above computations were repeated. The eccentricity is the mean value for the solar system satellites \citep{tarnopolski2}, and $\omega^2$ was decreased slightly (still characterizing a very oblate object) when compared to the value of Hyperion because in the employed method this parameter is treated as a small perturbation. The chaotic zone in case $\eta=0$, while much more narrow, is diffused in a large portion of the phase space (not shown). For values $\eta=1.7\!-\!2.2$, chaos is suppressed as spectacularly as previously demonstrated, in which case the ratio of the control term (for $\eta=1.7$) to the potential $\mathcal{V}$ is about 0.14, i.e. twice as small as when the parameters of Hyperion were employed.

\section{Discussion and conclusions}\label{sect5}

The Hamiltonian $\mathcal{H}$ in Eq.~(\ref{eq33}), yielding Eq.~(\ref{eq5}), was employed in Sect.~\ref{sect262} to construct a control term $\mathcal{F}$ that is able to suppress the chaotic behavior, which forms a new Hamiltonian $\tilde{\mathcal{H}}=\mathcal{H}+\eta\mathcal{F}$. The zeroth-order first term of the series from Eq.~(\ref{eq30}), i.e. the zeroth-order series expansion of Eq.~(\ref{eq31}), $\mathcal{F}_{2(0)}$, was constructed as described in Sect.~\ref{sect26}, resulting in Eq.~(\ref{eq34}). An additional multiplicative scaling term $\eta$ was introduced, and the resulting equation of motion in Eq.~(\ref{eq36}) was integrated with $\eta\in[0,9.9]$; $\eta=0$ corresponds to the lack of control term and served as a reference. The results represented in Fig.~\ref{fig14} show that the diffusion in the phase space cannot only be diminished, but when $\eta=3$ (meaning that the control term is an order of magnitde smaller than the potential $\mathcal{V}$) the motion becomes exactly periodic, which is confirmed with the power spectrum in Fig.~\ref{fig15}. With $\eta=2.2$, the motion becomes quasi-periodic with a very small diffusion in the phase space. Therefore, chaos is successfully suppressed.

While the model of planar rotation is in fact not applicable to Hyperion (used herein as a demonstration because, for its parameters, the phenomenon investigated herein is prominently visible), it should more adequately describe other oblate solar system satellites \citep{melnikov1}. This is indeed the case, as the diffusion of the chaotic trajectory, with parameters typical for a solar system asteroid, is suppressed to nearly periodic rotation. The control term is then only about 14\% of the potential. For even smaller $\omega^2$, which is a perturbative parameter, the suppression should be expected to be even more efficient.

\onecolumn

\appendix

\section{Beletskii equation}\label{appA}

From Fig.~\ref{fig1} it follows that $\theta=\alpha+f$; hence $\theta'=\alpha'+1$ and $\theta''=\alpha''$, where the prime denotes differentiation with respect to $f$. Substituting this into Eq.~(\ref{eq5}),
\begin{equation}
(1+e\cos f)\alpha''-2e\sin f (\alpha'+1)+\frac{\omega^2}{2}\sin 2\alpha=0,
\label{eqA1}
\end{equation}
which after defining $\delta=2\alpha$ and reordering becomes the \citet{beletskii3} equation, i.e.
\begin{equation}
(1+e\cos f)\delta''-2e\sin f\delta'+\omega^2\sin\delta=4e\sin f.
\label{eqA2}
\end{equation}

\section{Explicit formulae}\label{appB}

The potential from Eq.~(\ref{eq33}) in the form of Eq.~(\ref{eq28}) is as follows:
\begin{equation}
\begin{split}
\mathcal{V} = & -\frac{e\omega^2}{16}\exp[\i(-2\theta+f)]-\frac{e\omega^2}{8}\exp[\i(-2\theta+2f)]-\frac{e\omega^2}{16}\exp[\i(-2\theta+3f)]+\\
& -\frac{e\omega^2}{16}\exp[\i(2\theta-f)]-\frac{e\omega^2}{8}\exp[\i(2\theta-2f)]-\frac{e\omega^2}{16}\exp[\i(2\theta-3f)]
\end{split}
\label{eqB1}
.\end{equation}

The term $\Gamma \mathcal{V}$, i.e.
\begin{equation}
\begin{split}
\Gamma \mathcal{V} = & -\frac{1}{64 (e \cos f (e \cos f+2)-2 p+1) (e \cos f (e \cos f+2)-p+1) (3 e \cos f (e \cos f+2)-2 p+3)}\times\\
& \big[\omega ^2 (e \cos f+1)^3 \big(6 e^4 \sin (4 f-2 \theta )+e^4 \sin (6 f-2 \theta )+12 e^4 \sin 2 (f-\theta )-3 e^4 \sin 2 (f+\theta )+\\
& -10 e^4 \sin 2 \theta +9 e^3 \sin (5 f-2 \theta )-21 e^3 \sin (f+2 \theta )+30 e^2 \sin (4 f-2 \theta )+84 e^2 \sin 2 (f-\theta )+\\
& -12 e^2 p \sin (4 f-2 \theta )-32 e^2 p \sin 2 (f-\theta )+3 e \big(17 e^2-24 p+20\big) \sin (f-2 \theta )+\\
& +e\big(39 e^2-56 p+44\big) \sin (3 f-2 \theta )-54 e^2 \sin 2 \theta +20 e^2 p \sin 2 \theta +24 \sin 2 (f-\theta )+\\
& +32 p^2 \sin 2 (f-\theta )-64 p \sin 2 (f-\theta )\big)\big]
\end{split}
\label{eqB2}
\end{equation}
leads to the control term
\begin{equation}
\begin{split}
\mathcal{F}_2 = & -\frac{1}{256 (-2 p+e \cos f (e \cos f+2)+1)^2 (-p+e \cos f (e \cos f+2)+1)^2 (-2 p+3 e \cos f (e \cos f+2)+3)^2}\times\\
& \Big[(e \cos f \omega +\omega )^4 \sin 2 (\theta -f) \big(4 (-2 p+e \cos f (e \cos f+2)+1) (-p+e \cos f (e \cos f+2)+1)\times\\
& (-2 p+3 e \cos f (e \cos f+2)+3) \big(3 \sin (4 f-2 \theta) e^2+8 \sin 2 (f-\theta ) e^2-5 \sin 2 \theta e^2+18 \sin (f-2 \theta ) e+\\
& +14 \sin (3 f-2 \theta ) e-16 p \sin 2 (f-\theta )+16 \sin 2 (f-\theta )\big)-2 (-2 p+e \cos f (e \cos f+2)+1)\times\\
& (-p+e \cos f (e \cos f+2)+1) \big(6 \sin (4 f-2 \theta ) e^4+\sin (6 f-2 \theta ) e^4+12 \sin 2 (f-\theta ) e^4-10 \sin 2 \theta e^4+\\
& -3 \sin 2 (f+\theta ) e^4+9 \sin (5 f-2\theta ) e^3-21 \sin (f+2 \theta ) e^3-12 p \sin (4 f-2 \theta ) e^2+30 \sin (4 f-2 \theta ) e^2+\\
& -32 p \sin 2 (f-\theta ) e^2+84 \sin 2 (f-\theta ) e^2+20 p \sin 2 \theta e^2-54 \sin 2\theta e^2+3 \big(17 e^2-24 p+20\big)\times\\
& \sin (f-2 \theta ) e+\big(39 e^2-56 p+44\big) \sin (3 f-2 \theta ) e+32 p^2 \sin 2 (f-\theta )-64 p \sin 2 (f-\theta )+\\
& +24 \sin 2 (f-\theta)\big)-(-2 p+e \cos f (e \cos f+2)+1) (-2 p+3 e \cos f (e \cos f+2)+3)\times\\
& \big(6 \sin (4 f-2 \theta ) e^4+\sin (6 f-2 \theta ) e^4+12 \sin 2 (f-\theta ) e^4-10 \sin 2 \theta e^4-3\sin 2 (f+\theta ) e^4+\\
& +9 \sin (5 f-2 \theta ) e^3-21 \sin (f+2 \theta ) e^3-12 p \sin (4 f-2 \theta ) e^2+30 \sin (4 f-2 \theta ) e^2-32 p \sin 2 (f-\theta ) e^2+\\
& +84 \sin 2 (f-\theta ) e^2+20 p \sin 2 \theta e^2-54 \sin 2 \theta e^2+3 \big(17 e^2-24 p+20\big) \sin (f-2 \theta ) e+\\
& +\big(39 e^2-56 p+44\big) \sin (3 f-2 \theta ) e+32 p^2 \sin 2 (f-\theta )-64 p \sin 2(f-\theta )+24 \sin 2 (f-\theta )\big)+\\
& -2 (-p+e \cos f (e \cos f+2)+1) (-2 p+3 e \cos f (e \cos f+2)+3) \big(6 \sin (4 f-2 \theta ) e^4+\sin (6 f-2 \theta ) e^4+\\
& +12 \sin 2 (f-\theta) e^4-10 \sin (2 \theta ) e^4-3 \sin 2 (f+\theta ) e^4+9 \sin (5 f-2 \theta ) e^3-21 \sin (f+2 \theta ) e^3+\\
& -12 p \sin (4 f-2 \theta ) e^2+30 \sin (4 f-2 \theta ) e^2-32 p \sin 2 (f-\theta ) e^2+84 \sin 2 (f-\theta ) e^2+20 p \sin 2 \theta e^2+\\
& -54 \sin 2 \theta e^2+3 \big(17 e^2-24 p+20\big) \sin (f-2 \theta ) e+\big(39 e^2-56 p+44\big) \sin (3 f-2 \theta ) e+32 p^2 \sin 2 (f-\theta )+\\
& -64 p \sin 2 (f-\theta )+24 \sin 2 (f-\theta )\big)\big)\Big],
\end{split}
\label{eqB3}
\end{equation}
which with the substitution $p=\left(1+e\cos f\right)^2\theta'$ turns into
\begin{equation}
\begin{split}
\mathcal{F}_2 = & -\frac{\omega^4}{32 \big(1-2 \theta '\big)^2 \big(3-2 \theta '\big)^2 \big(\theta '-1\big)^2 (e \cos f+1)}\times\\
& \Big[\sin 2 (f-\theta ) \big(2 e \big(2 \theta'^2-5 \theta '+3\big)^2 \sin (f-2 \theta )+2 e \big(2 \theta'^2-3 \theta '+1\big)^2 \sin (3 f-2 \theta)+\\
& \big(4 \big(\theta '-2\big) \theta '+3\big)^2 \sin 2 (f-\theta )\big)\Big].
\end{split}
\label{eqB4}
\end{equation}
After a power series expansion in $\theta'$ to a zeroth order, Eq.~(\ref{eq34}) is obtained.

\end{document}